# Optimal Fully Electric Vehicle load balancing with an ADMM algorithm in Smartgrids

Andrea Mercurio, Alessandro Di Giorgio, Fabio Purificato
Department of Computer, Control and Management Engineering "Antonio Ruberti"
University of Rome "Sapienza", Via Ariosto 25, Rome, 00185, Italy
{mercurio, digiorgio}@dis.uniroma1.it

*Abstract* — **In this paper we present a system architecture and a suitable control methodology for the load balancing of Fully Electric Vehicles at Charging Station (CS). Within the proposed architecture, control methodologies allow to adapt Distributed Energy Resources (DER) generation profiles and active loads to ensure economic benefits to each actor. The key aspect is the organization in two levels of control: at local level a Load Area Controller (LAC) optimally calculates the FEV's charging sessions, while at higher level a Macro Load Area Aggregator (MLAA) provides DER with energy production profiles, and LACs with energy withdrawal profiles. Proposed control methodologies involve the solution of a Walrasian market equilibrium and the design of a distributed algorithm.**

## I. Introduction

Electric Vehicle (EV) technology has been attracting growing interest in recent years. By 2035, it is expected that EV technology will play around 40% share of global car market [1]. This is fundamentally due to its capacity of being an environmental friendly transportation mean and to the fact that its integration with the power grid infrastructure is a non-trivial issue due to the necessity of handling an increased load with impacts on the efficiency of the distribution network [2],[3]. Smartgrid research studies show that EVs can play a significant role in the Active Demand (AD), having a potential of participating to the definition of Demand Side Management (DSM) services [4],[5],[6] and in the regulation of the distribution network [7]. The FP7 funded Mobincity research project [8] is to enable wide deployment of Full-Electric Vehicles (FEV) as mass market product in cities. With this regard, Mobincity aims to optimise both energy charging and discharging processes and to increase energy efficiency, allowing a seamless Vehicle To Grid (V2G) integration by exploiting the potentialities of releasing DSM services based on FEVs' controllable storage devices. Effectively exploiting such potential would allow FEVs to significantly contribute to the Smartgrid development. The aim of this work is to present the control structure adopted for the integration of Mobincity's architecture into the energy infrastructure grid and a methodology for load balancing. In this work the V2G interaction is assumed to be mono-directional, being this the most widespread and likely solution for FEVs integration into the power grid, at least for the first years. The proposed approach falls into the field of Active Network Management (ANM) as exposed in the taxonomy proposed in [9].

This work is partially supported by the European Union FP7 ICT-GC MOBINCITY project, grant agreement no. 314328.

## II. System Architecture and Related Work

As common in Smartgrid projects, such as FP7 funded ADDRESS project [10], Mobincity adopts a two level control approach for the load management (Figure 1), being the load constituted by FEVs at Charging Stations (CS). In order to assure a scalable methodology, load aggregates are divided into Load Areas (LAs) and Macro Load Area (MLAs) according to [11] and [12]. At lower level a Load Area Controller (LAC) optimally schedules and manages the charging sessions of the FEVs on CSs insisting on a single Load Area (LA), considering FEV user's needs and local microgeneration. Solutions to the LAC problem are formulated and discussed in [13] and [14]. At higher level a Macro Load Area Aggregator (MLAA) balances, in advance with respect to a specific time horizon, the aggregated load, the Distributed Energy Resources (DER) and the grid energy withdrawal, and, under requests of authorised actors (System Operators, Generation Unit Operators, Retailers), composes and releases in real-time AD products, the latters being defined in [11] and [12]. This paper presents a method for the balancing problem of the MLAA, which provides reference consumption to LACs, Retailer and generation profiles to DER Operators on the specific time horizon; all actors behave like an energy balancing group respecting the overall consumption programme. Future works will face the AD product composition and release processes. The balancing is realised through an Alternate Direction Method of Multipliers (ADMM) algorithm [15], in which market actors (LACs, DER operator, Retailers) interact with the MLAA that plays the role of the price caller. A Walrasian competitive equilibrium [16] is therefore realised within the Macro Load Area. The LAC and MLAA are assumed to be components of the Electric Vehicle Support Equipment (EVSE), that is the set of CSs and related information system owned by an EVSE operator, which offers the EV charging services. Such architecture, implemented with Web Services, could be integrated into the Sustainable Energy Microsystem (SEM) of Smartgrids [17]. Communication means between actors are likely to be implemented over the next generation internet [18]. In research studies concerning the V2G integration of EV many approaches have been proposed. Vandael *et al.* [4] proposed an interesting three-step approach, based on demand and supply functions, which blends both centralized and decentralized methods for scheduling the FEV charging. Sortomme *et al.* [5] simulated a number of centralized algorithms for an aggregating entity that both procures energy for the FEVs and provides load regulation as a form of DSM, providing benefits to customers, profits to aggregator and regulation services to utilities. The work described in [7] proposes a local algorithm to minimize the time of charge of an EV whilst assuring Distribution Network constraints. Metaheuristics methods, such as Tabu Search [19], have also been proposed. The two level approach of Mobincity separates the problem of optimally allocating the FEV charging sessions, being this a task accomplished in a centralized way in the LACs, whilst the MLAA takes care of load balancing and AD product composition, adopting a de-centralized approach, thus enforcing scalability. In place of demand and supply functions, utility functions for consumers and cost functions for producers are used instead. LACs and other actors do not need to communicate utility functions and cost functions, thus assuring privacy. The proposed methodology intends to ameliorate the approach proposed in [20] and [21]. In fact, a better definition of the consumer utility function is proposed and the ADMM algorithm assures better convergence properties due to the augmented lagrangian formulation and only requires convexity of cost and utility functions. Furthermore ADMM allows a de-centralised implementation meant to overcome the scalability and memory problems highlighted in [22].

## III. MLAA BALANCING PROBLEM FORMALISATION

The MLAA deals with a number of actors present on the MLA: LACs, DER Operators, Retailers. DER operators might include both gas fueled micro-turbines, Combined Heat and Power (CHP) units or renewables. Similarly to [20], LACs' need for energy can be modeled by utility functions [23], the power producers are represented by cost functions and the load balancing is realized over a time horizon of $T$ time slots. For the list of symbols please refer to TABLE I. Utility and cost functions must be convex and continuous. The MLAA maximization problem is therefore:

$$max \sum_{t\in T}[\sum_{r\in\xi_D} U_r(x_r(t)) - \sum_{l\in\xi_G} C_l(c_l(t))] \qquad (1)$$

subject to the balancing constraints over the MLA:

$$\sum_{l\in\xi_G} c_l(t) - \sum_{r\in\xi_D} x_r(t) = 0\, \forall t \in T \qquad (2)$$

$$m_r(t) \leq x_r(t) \leq M_r(t) \forall t \in T$$

$$c_{lmin}(t) \leq c_l(t) \leq c_{lmax}(t) \forall t \in T$$

Equation (2) implies that there will be a unique price for the MLA in each time slot, differently from [20][21]. The second and third constraint are "local constraints" that assure that LACs and generators do not violate operative limits. The utility function for the LAC is:

$$U_r(x_r) = U^{MAX}[1 - e^{\frac{-x_r(t)}{Kx_{pr,r}(t)}}] \qquad (3)$$

The parameter $K$ is adjusted to set the error for which $U_r(x_r) \to U^{MAX}$ when $x_r \to x_{pr}$. Assuming $1\%$ error is desired $K$=0,217.

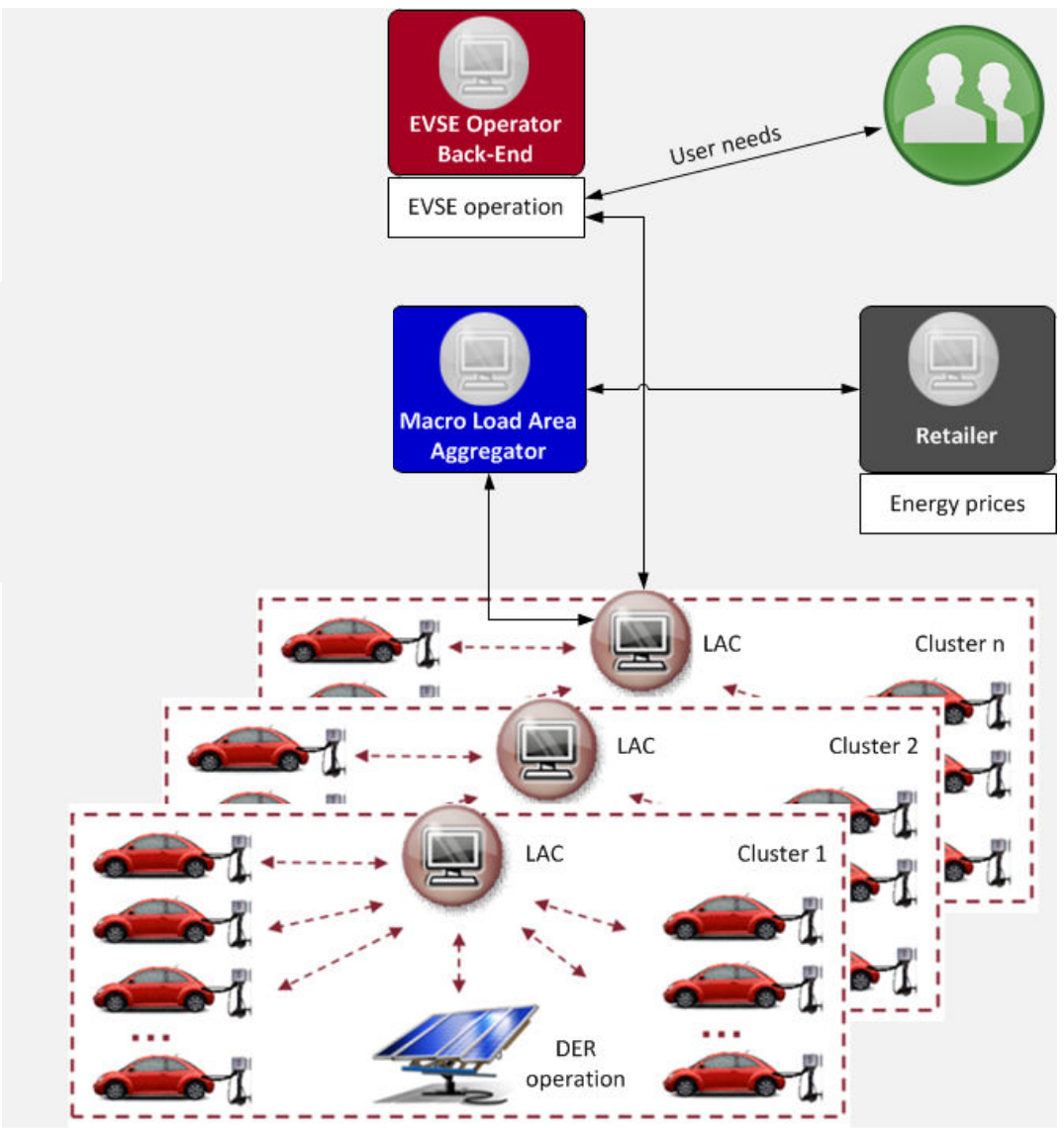

Figure 1. Mobincity Two levels control

TABLE I: List of symbols

| Symbol | Meaning |
|---|---|
| $r$ | Generic LAC index |
| $l$ | Generic generation unit index |
| $T$ | Time horizon |
| $\xi_G$ | Set of generation units |
| $\xi_D$ | Set of LAC |
| $x_r(t)$ | Power consumption of LAC $r$ in time slot $t$ |
| $m_{rl}(t); M_{rl}(t)$ | Min and Max power consumption for LAC $r$ |
| $c_l(t)$ | Power production of generation unit $l$ in time slot $t$ |
| $c_{min}(t); c_{max}(t)$ | Min. and Max. power generation for plant $l$ |
| $U_r(.)$ | Utility function of LAC $r$ |
| $C_l(.)$ | Cost function of generation unit $l$ |
| $K$ | Utility function parameter |
| $U_r^{MAX}$ | Maximum utility of LAC $r$ |
| $x_{pr,r}(t)$ | Desired power consumption of LAC $r$ in the $t$ time slot |
| $pv$ | Generic Photovoltaic plant index |
| $tpp$ | Generic fuel fed generation unit index |
| $N$ | Number of actors in the MLA |
| $y_i$ | Generic power absorption/production of the $i^{th}$ actor |
| $f_i(y_i)$ | Generic cost/utility function of the $i^{th}$ actor |
| $\bar{y} = \sum_{i=1}^{N} \frac{y_i}{N}$ | Average power absorption/production of all actors |
| $PV = \xi_{PV}$ | PV set |
| $TPP = \xi_{TPP}$ | TPP set |
| $GRID$ | Power grid |
| $c_{tpp}(t)$ | Power injection of a generic TPP in time slot $t$ |
| $c_{pv}(t)$ | Power injection of a generic PV in time slot $t$ |
| $c_{GRID}(t)$ | Power injection of a power grid in time slot $t$ |
| $\alpha_{tpp}$ | Parameter of cost function of a generic TPP |
| $\beta_{tpp}$ | Parameter of cost function of a generic TPP |
| $\gamma_{tpp}$ | Parameter of cost function of a generic TPP |
| $\Delta t$ | Time slot period |
| $\kappa_{GRID}(t)$ | Power grid cost in time slot $t$ |
| $\lambda(t)$ | MLA price for energy in time slot $t$ |
| $\tilde{\lambda}(t)$ | Forecasted MLA price for energy in time slot $t$ |
| $\rho(t)$ | Penalty parameter of the augmented Lagrangian |
| $k$ | Iteration step of the ADMM solution method |

$U^{MAX}$ can be then fixed to assure that $x_{pr,r}(t)$ is acquired, assuming the forecasted MLA price in the time slot $t$ is $\tilde{\lambda}$. This can be achieved by solving the following:

$$\lim_{x_r \to x_{pr}} \left\{ \frac{d}{dx_r} U_{max}(1 - e^{-\frac{x_r}{Kx_{pr}}}) - \tilde{\lambda} x_r \right\} = 0 \tag{4}$$

which gives the optimal value:

$$U_{max}{}^* = \tilde{\lambda} K x_{pr} e^{\frac{1}{K}} \tag{5}$$

In case of errors in the prediction of the MLA price the quantity of energy procured by each LAC would then vary according to:

$$\frac{x_r}{x_{pr}} = 1 + \mathrm{K} * \ln\frac{\tilde{\lambda}}{\lambda} \tag{6}$$

that is found by substitution of (5) into (4). The parameter $K$ is therefore a regulator of LAC's demand sensibility to price. This formulation of utility functions assures that, under predicted conditions, the desired quantity of energy is acquired, allowing, at the same time, elasticity to price value. Generation cost functions are:

$$C_{tpp}\left(c_{tpp}(t)\right) = \alpha_{tpp}(t) c_{tpp}^2(t) + \beta_{tpp}(t) c_{tpp}(t) + \gamma_{tpp}(t) \tag{7}$$

$$C_{pv}\left(c_{pv}(t)\right) = 0 \tag{8}$$

$$C_{GRID}\left(c_{GRID}(t)\right) = \kappa_{GRID}(t) c_{GRID}(t) \tag{9}$$

assuming: a quadratic cost function [26] for fuel fed generators (TPP), no generation costs for photovoltaic and linear costs for the power grid. More complex constraints on generators can be considered [4], affecting the definition of the solving method. In this work such constraints are not considered for ease of discussion.

## IV. SOLVING METHOD

Problem (1) is a convex optimization problem that can be solved with a large variety of methods [24][25]. The ADMM [15] is attractive because it offers the scalability advantages of a distributed solution and the convergence properties of the Augmented Lagrangian methods. The Augmented Lagrangian formulation of the original problem is:

$$\begin{aligned} max_{xc}(L_{\rho\lambda}(x,c)) = max_{xc} \sum_{t \in T} \Bigg\{ & \sum_{r \in \xi_D} U_r(x_r(t)) - \sum_{l \in \xi_G} C_l(c_l(t)) \\ & + \lambda(t) \left[ \sum_{l \in \xi_G} c_l(t) - \sum_{r \in \xi_D} x_r(t) \right] \\ & + \frac{\rho(t)}{2} \left\| \sum_{l \in \xi_G} c_l(t) - \sum_{r \in \xi_D} x_r(t) \right\|_2^2 \Bigg\} \end{aligned} \tag{10}$$

While the solution of the dual problem is:

$$min_\lambda \phi(\lambda) = min_\lambda max_{xc}(L_{\rho\lambda}(x,c)) \tag{11}$$

Problems (10) and (11) cannot be decomposed due to the presence of a quadratic term, but the ADMM method allows a technique to overcome the issue. Problem (1)(2), apart for the local constraints, can be written in the generic form:

$$\max \sum_{i=1}^{N} f_i(y_i) \tag{12}$$

$$s.t. \sum_{i=1}^{N} y_i = 0$$

As shown in [15], this problem can be solved, for each time slot $t$ , in a distributed manner through an ADMM algorithm as follows:

$$y_i^{k+1} = argmax_{y_i}\left( f_i(y_i) + \lambda_i^k y_i + \frac{\rho^k}{2} \left\| \left( y_i - (y_i^k - \bar{y}^k) \right) \right\|_2^2 \right) \quad i = 1, \dots, N \tag{13}$$

$$\bar{y}^{k+1} = \frac{1}{N} \sum_{i=1}^{N} y_i^{k+1} \tag{14}$$

$$I\lambda^{k+1} = I\lambda^k + \rho^k I \bar{y}^{k+1} \tag{15}$$

being the primal and dual residuals:

$$r^{k+1} = I\bar{y}^{k+1} = I \frac{1}{N} \sum_{i=1}^{N} y_i^{k+1} \tag{16}$$

$$\begin{aligned} s^{k+1} = \rho^k \Big( (y_1^{k+1} - y_1^k) - (\bar{y}^{k+1} - \bar{y}^k); \dots; (y_N^{k+1} - y_N^k) \\ - (\bar{y}^{k+1} - \bar{y}^k) \Big) = \rho^k (\Delta y - I \Delta \bar{y}) \end{aligned} \tag{17}$$

Where $I$ is the unitary vector. Relation (15) states that there are $N$ lagrangian multipliers which are all equal, this is relevant for the stopping criteria (21). Relations (13) to (17) are found by manipulating (12) in the form:

$$\max \sum_{i=1}^{N} f_i(y_i) - g(z)$$

$$s.t.: \ y - z = 0$$

Where g(z) is the indicator function of the convex set $C$:

$$C = \left( y \in R \,\middle|\, \sum_{i=1}^{N} y_i = 0 \right)$$

Then the generic convex ADMM problem is solved:

$$y^{k+1} = argmax_y \left( \sum_{i=1}^{N} f_i(y_i) + \frac{\rho^k}{2} \left\| \left( y + \frac{\lambda^k}{\rho^k} + z^k \right) \right\|_2^2 \right)$$

$$z^{k+1} = \prod\nolimits_C \left( y^{k+1} + \frac{\lambda^k}{\rho^k} \right)$$

$$\lambda^{k+1} = \lambda^k + \rho^k (y^{k+1} - z^{k+1})$$

Where $\pi_c$ is the projection of $y^{k+1} + \frac{\lambda^k}{\rho^k}$ on the set $C$. By solving the above relations, (13) to (17) are found. The dual residual is found as follows: because the $y_i^{k+1}$ minimises, by definition, the Lagrangian

$$\begin{aligned} \textstyle\sum_{w \neq i}^{N} f_w(y_w^k) + f_i(y_i) + \lambda_i^k \left( y_i - (y_i^k - \bar{y}^k) \right) + \\ + \frac{\rho^k}{2} \left\| \left( y_i - (y_i^k - \bar{y}^k) \right) \right\|_2^2 \end{aligned} \tag{18}$$

we therefore have

$$\begin{aligned} 0 &= \nabla f_i(y_i^{k+1}) + \lambda_i^k + \rho^k \left( y_i^{k+1} - (y_i^k - \bar{y}^k) \right) = \\ &= \nabla f_i(y_i^{k+1}) + \lambda_i^k + \rho^k \bar{y}^k - \rho^k r^{k+1} + \rho^k \left( y_i^{k+1} - (y_i^k - \bar{y}^k) \right) = \\ &= \nabla f_i(y_i^{k+1}) + \lambda_i^{k+1} - \rho^k \bar{y}^k + \rho^k \left( y_i^{k+1} - (y_i^k - \bar{y}^k) \right) \end{aligned}$$

Considering that, for the dual feasibility condition

$$\nabla f_i(y_i^{k+1}) + \lambda_i^{k+1} = 0$$

than the dual residual for the i^th variable is:

$$s_i^{k+1} = \rho^k \left( (y_i^{k+1} - y_i^k) - (\bar{y}^{k+1} - \bar{y}^k) \right) \tag{19}$$

which justifies (17). Being the ADMM a dual-primal technique, the stopping criteria is based on the primal and dual problem residuals as follows [15]:

$$\|r^k\|_2 < \epsilon_{pri} = \sqrt{N}\epsilon_{abs} + \epsilon_{rel} max(\|y_i\|_2) \tag{20}$$

$$\|s^k\|_2 < \epsilon_{dual} = \sqrt{N}\epsilon_{abs} + \epsilon_{rel} \left\| \sum_{i=1}^{N} \lambda_i^k \right\|_2 \tag{21}$$

The following $\rho$ adjustment criteria applies:

$$\rho^{k+1} = \begin{cases} 2\rho^k \; if \; \|r^{k+1}\|_2 > 10 \; \|s^{k+1}\|_2 \\ \dfrac{\rho^k}{2} \; if \; \|r^{k+1}\|_2 < \dfrac{1}{10} \| \, s^{k+1}\|_2 \\ \rho^k \quad\quad otherwise \end{cases} \tag{22}$$

that is, at each step $k$, each LAC and power generation operator solves the primal problems (13) assuring that the local constraints hold, then communicates the consumption and generation values to the MLAA. The MLAA, at each iteration step, computes (14) to (17) and (20) to (22), then broadcasts (14), (15) and (22) to LACs and generators. This iteration ends when the stopping criteria is met. This is the Walrasian *tatônnement* method to reach market equilibrium [16]. Privacy is guaranteed as LACs and generators just exchange consumption and production quantities with the MLAA and they receive the MLA energy price and the penalty terms $\rho$, $\bar{y}$. The quadratic term for each primal problem represents an incentive term adjusted from the MLAA to attain the MLA balance. Convergence is assured as functions (3),(7) and (9) are closed, proper and convex, while assuming that the non-augmented lagrangian of (10) has a saddle point is not restrictive [15].

## V. RESULTS

The computational routines to solve the problem have been implemented with MatLab®. The simulation scenario is summarized in TABLE II.

TABLE II. MLA Simulation Scenario

| | | | | |
|---|---|---|---|---|
| **N° LACs** | **20** | Max absorption each LAC 200 kW | | |
| **DER** | **3** | **GRID cost function** | **TPP cost function** | **PV cost function** |
| | | $9.87 \left[\frac{€cent}{kWh}\right]$ 8:00 to 18:00 $18.21 \left[\frac{€cent}{kWh}\right]$ 18:00 to 8:00 Max 2MW | $0.02 * c_2^2(t) + 11.5 * c_2(t)$ Max 1MW Min 200kW | 0 [€cent/kWh] Max 1MW |

The PV power production has been assumed to be the same as Figure 2. The TPP has been assumed to have a minimum power to dispatch, as it were a cogeneration plant that cannot be shut down for conditioning/heating reasons. The simulation horizon is 24 hours divided in 24 time slots. The energy cost of the power grid is a typical bi-hourly tariff. The stopping criteria used is $\epsilon_{abs} = 10^{-4}$ and $\epsilon_{rel} = 10^{-5}$. LACs' desired power consumption profiles are reported in Figure 3. The desired power consumptions of LACs are uncorrelated. Power generation from TPP and power withdrawal from the grid are reported in TABLE II. Results are presented from Figure 3 to Figure 7. If the energy price predictions are sharp, the each LAC's actual consumption is affected by a small error (Figure 5) with respect to the desired power profile Figure 3. Figure 7 shows that the proposed algorithm converges in around 40 iterations.

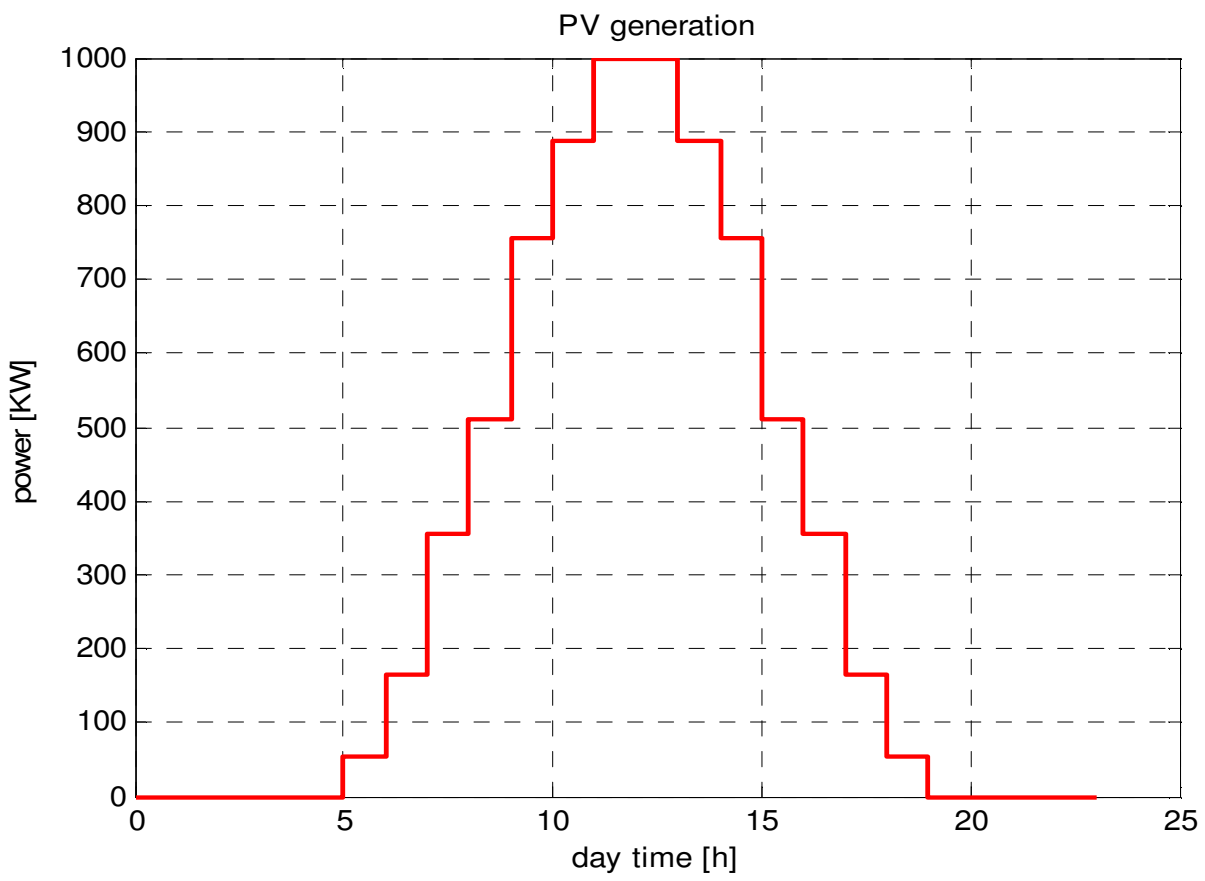

Figure 2. Photovoltaic generation profile.

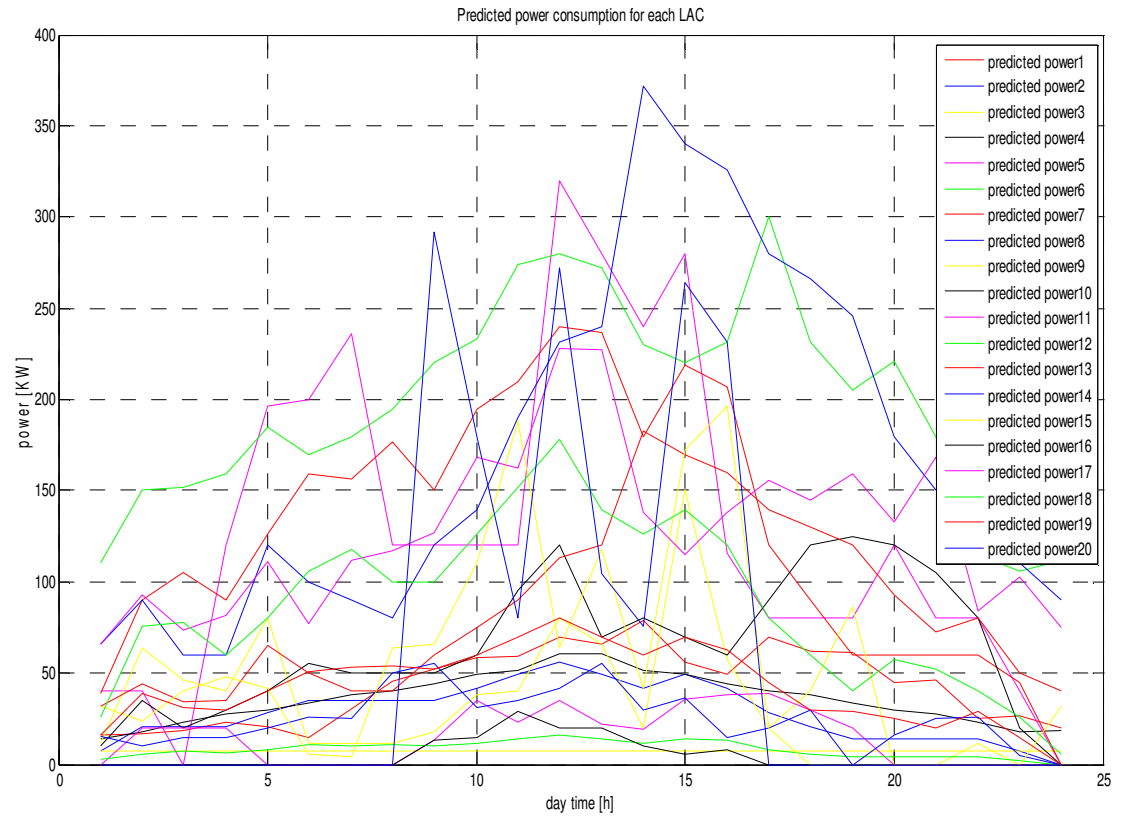

Figure 3. LACs desired power profile, simulation #1

As it can be seen, the algorithm dispatches the most economic unit first. TPP is at minimum when the GRID is off peak-price, then it is at maximum when GRID is in peak-price. The presence of PV provides beneficial effects to MLA prices, in particular when the power withdrawal from the GRID is nulled, such as from 9 a.m. to 10 a.m. The priority of despatchment of the most economic units is better emphasized by running a similar simulation using the same desired consumption as reported in Figure 8. The resulting power consumption and withdrawals are reported in Figure 9 and MLA's energy price in Figure 10. It is evident that a reduction of withdrawals from GRID and TPP corresponds to an increment of PV's injection. When the withdrawal from the GRID is nulled, during peak-price hours, the price falls. It is interesting to note that, in both simulations, the MLA energy price is determined by the most expensive power generator over the MLA, being it the TPP or the GRID. That implies extra incomes for the PV and for the TPP each time the MLA procures energy from the GRID, and vice-versa. The MLA price mechanism, therefore, dispatches the cheaper units before, guaranteeing them better incomes when the most expensive units are used.

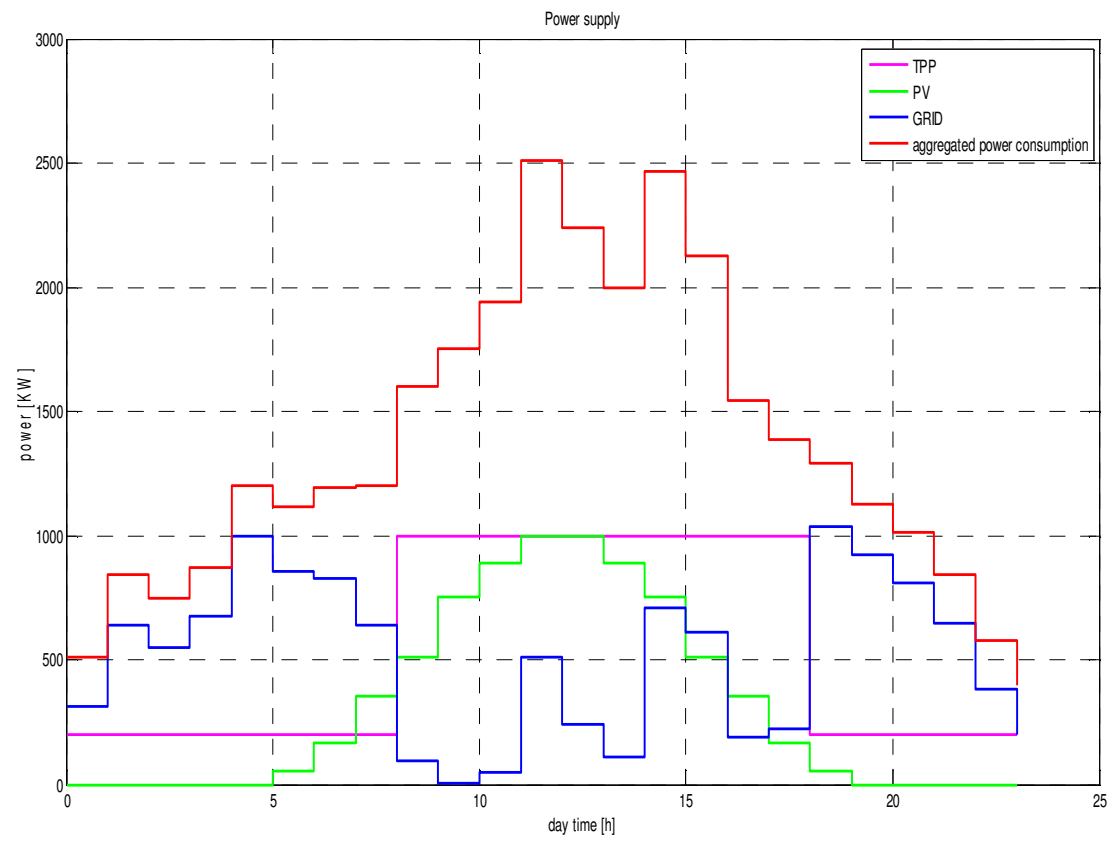

Figure 4. MLA power generation and Consumption simulation #1

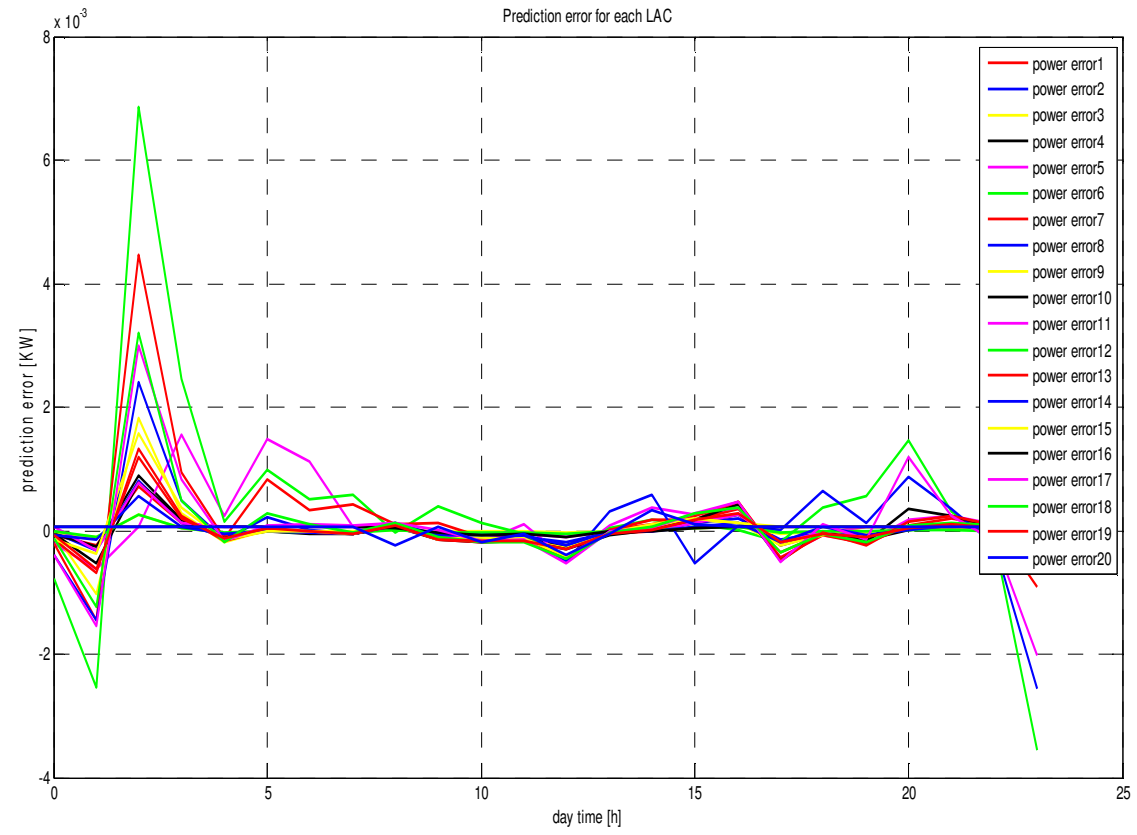

Figure 5. Single LACs power consumption error simulation #1

TABLE III, in fact, shows the average energy bill at the end of the day for a single LAC with and without DER, for simulation #1. As it can be seen there is little benefit, as GRID withdrawals are nulled in one time-slot only. The algorithm here proposed solves some of the shortfalls highlighted in [20], namely: the independence of energy withdrawals from energy resources and the high prices for PV energy when it is not producing. This is due to the better formulation of the utility functions and the MLA price mechanism. Moreover, the ADMM method proved to have better convergence performances. Nevertheless, the MLA price mechanism does not provide, itself, direct economic benefits to FEVs. This depends on the MLA FEV demand, DER production in the MLA, and the GRID costs.

## VI. Conclusions

This paper presented an architecture for the control of FEVs' load and a distributed algorithm for the balancing of load, DG and grid power withdrawal. Preliminary simulations confirm the capability of allowing the integration of Distributed Generation and Renewable Generation, which are fully dispatched locally.

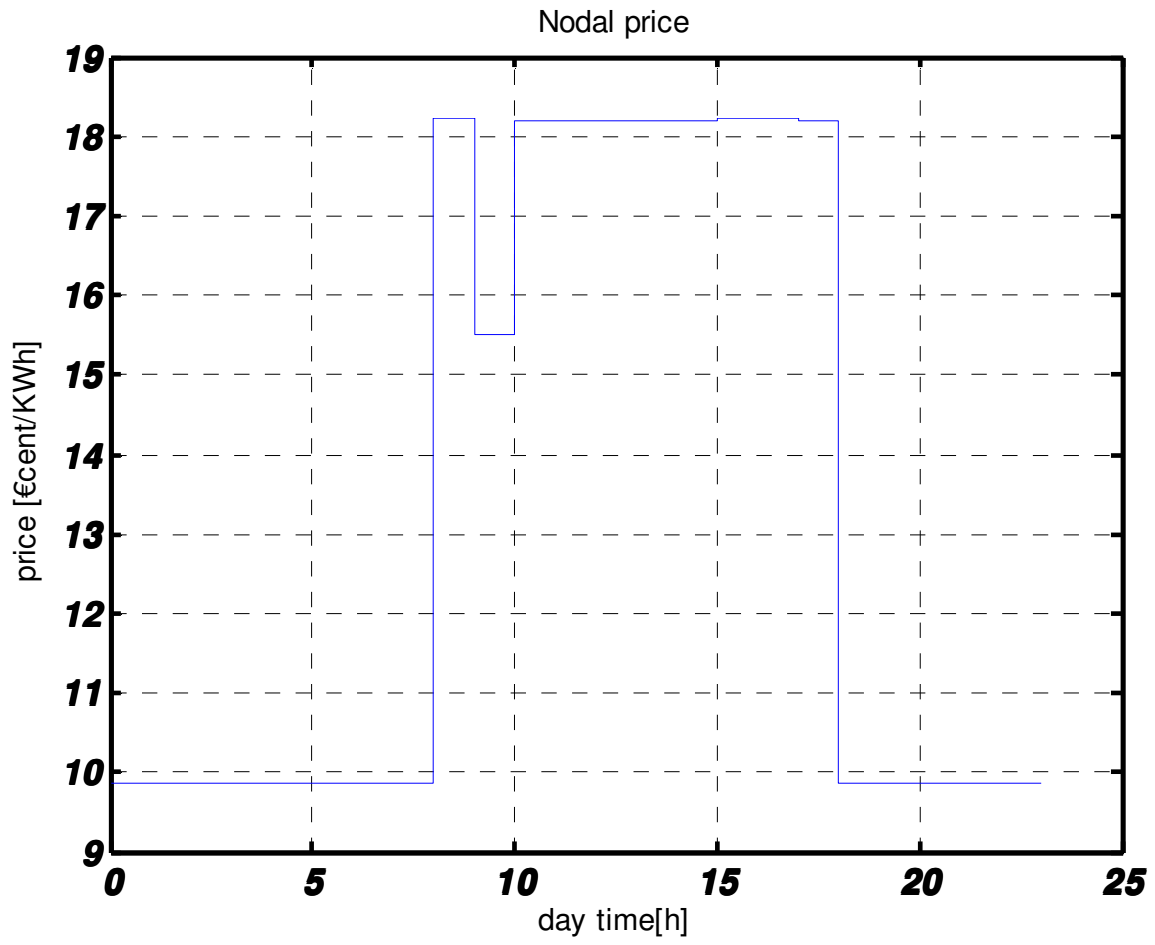

Figure 6. MLA energy price simulation #1

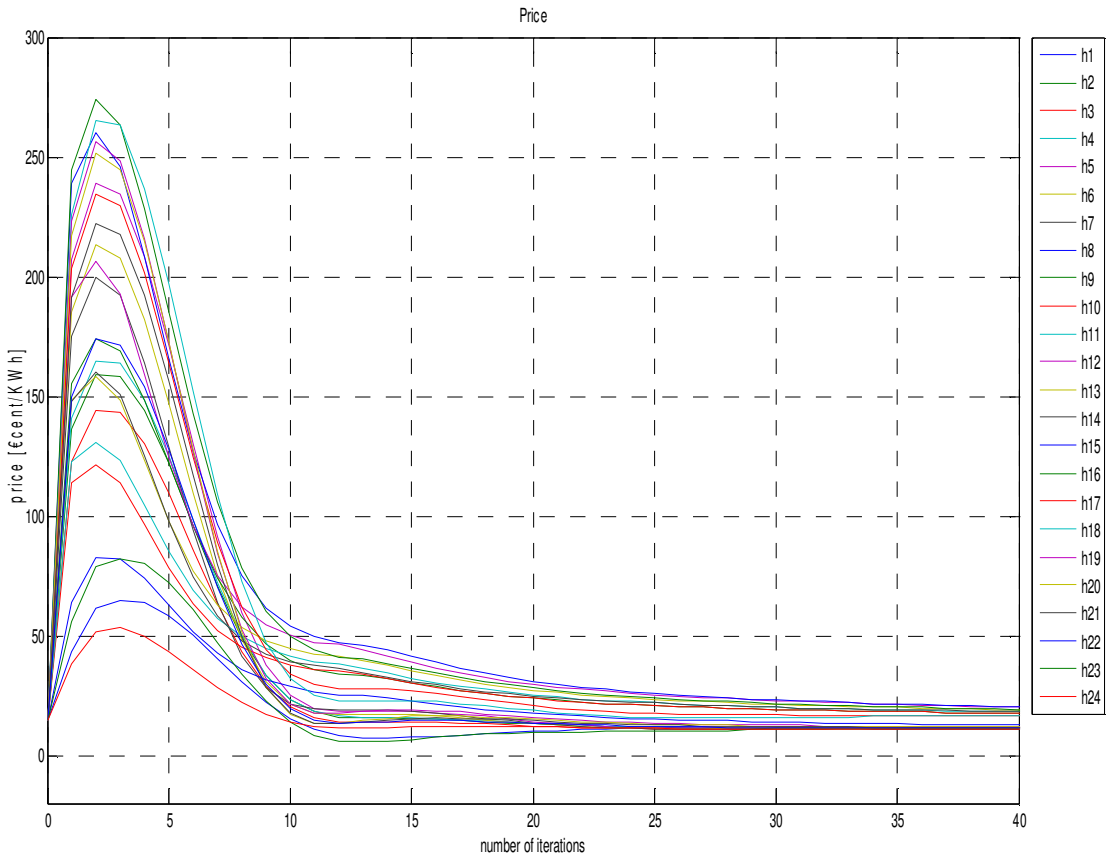

Figure 7. MLA price evolution simulation #1

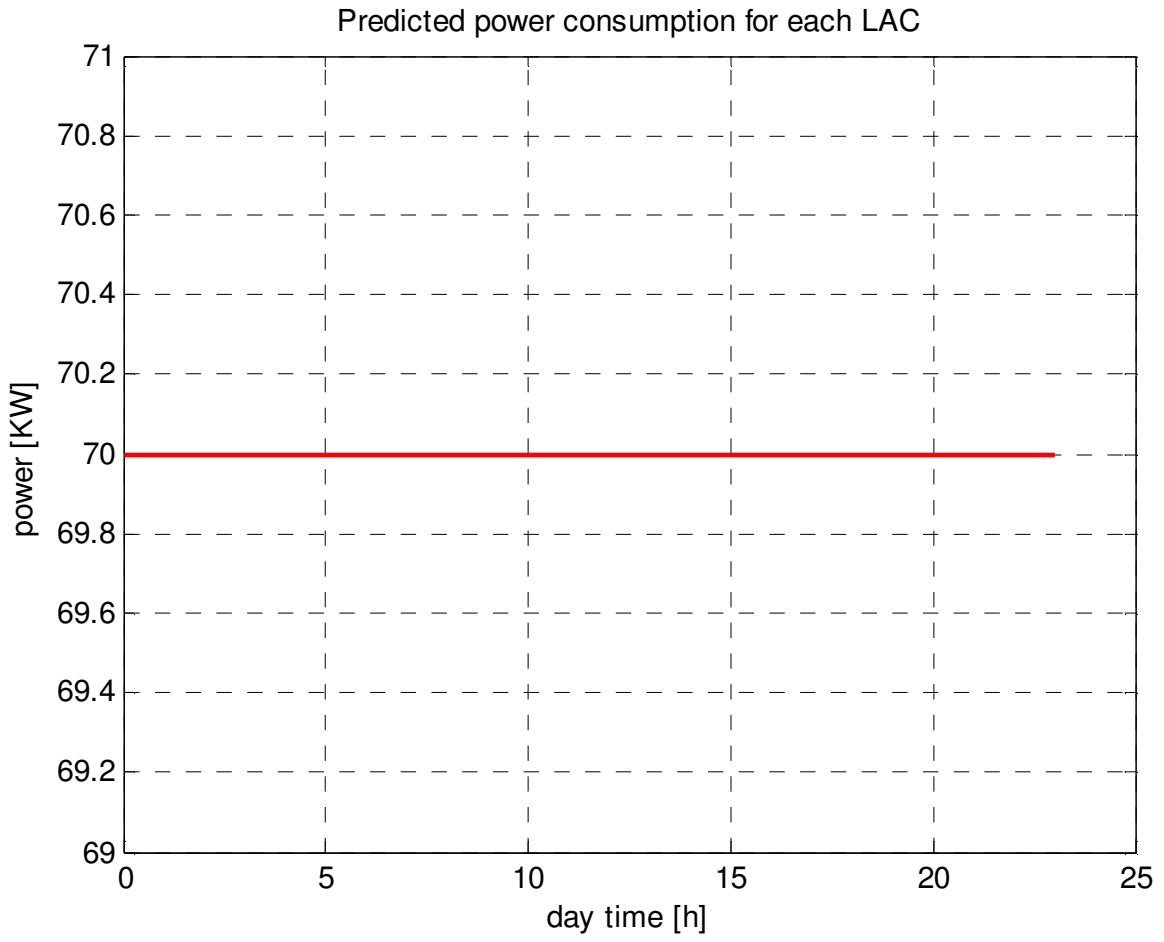

Figure 8. LACs desired power profile simulation #2

MLA's energy price is defined by the costs of the most expensive unit. Therefore, economic benefits to FEVs, with respect to a DER absence, are not assured, as MLA's energy price falls only when the most expensive resources are not used.

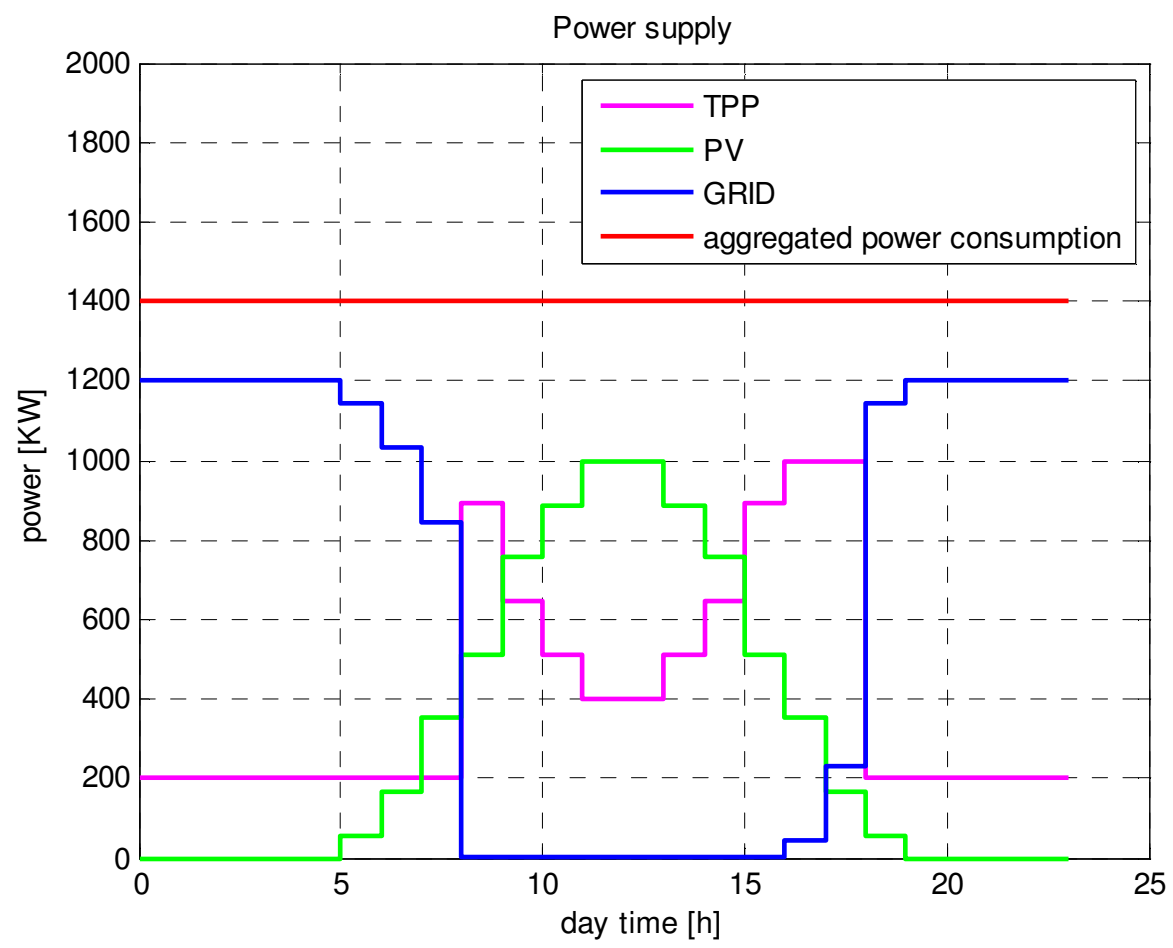

Figure 9. MLA power generation and Consumption simulation #2

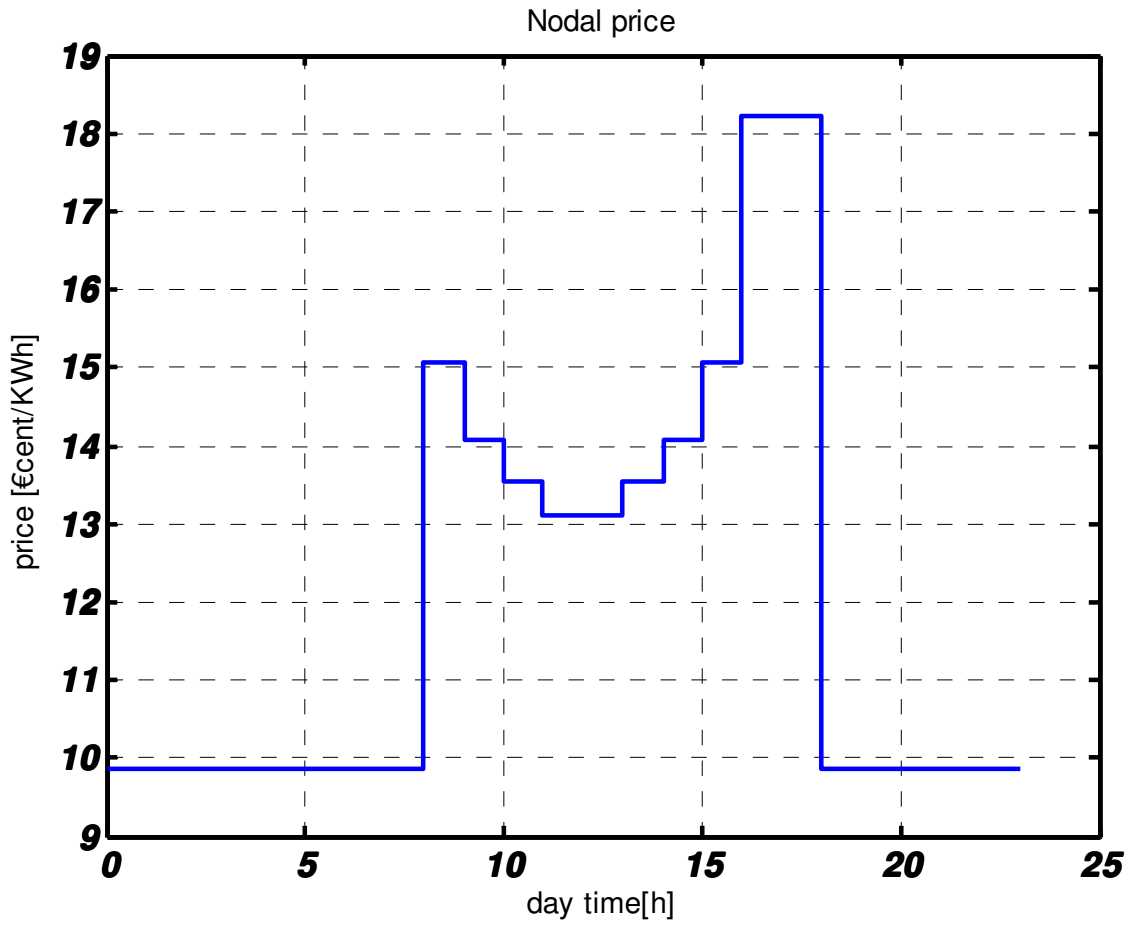

Figure 10. MLA energy price simulation #2

TABLE III. Energy Savings

| | DER presence | Power Grid Only |
|---|---|---|
| Total bill for a LAC [€/day] | 239.5 | 241.97 |

The algorithm proposed solves some shortfalls of previously proposed solutions, while keeping privacy advantages. Currently authors are working to extend this approach to include network constraints and to define the AD product composition and release processes.